\documentstyle[12pt]{article}
\font\tenrm=cmr10
\font\sma=cmr10 at 10truept

\def\al{\alpha}

\def\ga{\gamma}
\def\de{\delta}

\def\th{\theta}

\def\ka{\kappa}

\def\ph{\phi}

\def\ps{\psi}

\def\Ga{\Gamma}

\def\Ps{\Psi}

\def\fr#1#2{{{#1} \over {#2}}}

\def\vev#1{\langle {#1}\rangle}

\def\half{{\textstyle{1\over 2}}}
\def\frac#1#2{{\textstyle{{#1}\over {#2}}}}

\def\lsim{\mathrel{\rlap{\lower4pt\hbox{\hskip1pt$\sim$}}
    \raise1pt\hbox{$<$}}}
\def\gsim{\mathrel{\rlap{\lower4pt\hbox{\hskip1pt$\sim$}}
    \raise1pt\hbox{$>$}}}
\def\sqr#1#2{{\vcenter{\vbox{\hrule height.#2pt
         \hbox{\vrule width.#2pt height#1pt \kern#1pt
         \vrule width.#2pt}
         \hrule height.#2pt}}}}

\newcommand{\beq}{\begin{equation}}
\newcommand{\eeq}{\end{equation}}
\newcommand{\bea}{\begin{eqnarray}}
\newcommand{\eea}{\end{eqnarray}}
\newcommand{\rf}[1]{(\ref{#1})}

\renewenvironment{thebibliography}[1]
 { \rm
   \begin{list}{\arabic{enumi}.}
    {\usecounter{enumi} \setlength{\parsep}{0pt}
     \setlength{\itemsep}{3pt} \settowidth{\labelwidth}{#1.}
     \sloppy
    }}{\end{list}}

\begin{document}
\titlepage

\begin{flushright}
{COLBY 97-07\\}
{IUHET 371\\}
{September 1997\\}
\end{flushright}
\vglue 1cm
	    
\begin{center}
{{\bf REVIVALS OF RYDBERG WAVE PACKETS \footnote[1]{\tenrm 
Talk presented by R.B. at the 
VIII International Conference on Symmetry Methods
in Physics, Dubna, Russia, July, 1997}
\\}
\vglue 1.0cm
{Robert Bluhm$^a$, 
V. Alan Kosteleck\'y$^b$,
and B. Tudose$^c$
\\} 
\bigskip
{\it $^a$Physics Department, Colby College\\}
\medskip
{\it Waterville, ME 04901, U.S.A.\\}
\vglue 0.3cm
{\it $^b$Physics Department, Indiana University\\}
\medskip
{\it Bloomington, IN 47405, U.S.A.\\}
\vglue 0.3cm
{\it $^c$Physics Department, Northwestern University\\}
\medskip
{\it Evanston, IL 60208, U.S.A.\\}

\bigskip

\vglue 0.8cm
}
\vglue 0.3cm

\end{center}

{\rightskip=3pc\leftskip=3pc\noindent
We examine the revival structure of Rydberg wave packets. 
These wave packets exhibit initial classical periodic motion
followed by a sequence of collapse, 
fractional/full revivals, 
and fractional/full superrevivals.
The effects of quantum defects on wave packets in
alkali-metal atoms and a squeezed-state description
of the initial wave packets are also described.
We then examine the revival structure of Rydberg wave packets 
in the presence of an external electric field, 
i.e., the revival structure of Stark wave packets. 
These wave packets have energies that depend on two quantum numbers
and exhibit new types of interference behavior.
}

\vfill
\newpage

\baselineskip=12pt

\begin{center}
{{\bf REVIVALS OF RYDBERG WAVE PACKETS \footnote[1]{\tenrm 
Talk presented by R.B. at the 
VIII International Conference on Symmetry Methods
in Physics, Dubna, Russia, July, 1997}
\\}
\vglue 1.0cm
{Robert Bluhm$^a$, 
V. Alan Kosteleck\'y$^b$,
and B. Tudose$^c$
\\} 
\bigskip
{\it $^a$Physics Department, Colby College\\}
\medskip
{\it Waterville, ME 04901, U.S.A.\\}
\vglue 0.3cm
{\it $^b$Physics Department, Indiana University\\}
\medskip
{\it Bloomington, IN 47405, U.S.A.\\}
\vglue 0.3cm
{\it $^c$Physics Department, Northwestern University\\}
\medskip
{\it Evanston, IL 60208, U.S.A.\\}

\vglue 0.8cm
}
\vglue 0.3cm

\end{center}

{\rightskip=3pc\leftskip=3pc\noindent\sma
We examine the revival structure of Rydberg wave packets. 
These wave packets exhibit initial classical periodic motion
followed by a sequence of collapse, 
fractional/full revivals, 
and fractional/full superrevivals.
The effects of quantum defects on wave packets in
alkali-metal atoms and a squeezed-state description
of the initial wave packets are also described.
We then examine the revival structure of Rydberg wave packets 
in the presence of an external electric field, 
i.e., the revival structure of Stark wave packets. 
These wave packets have energies that depend on two quantum numbers
and exhibit new types of interference behavior.
}

\baselineskip=16pt

\vglue 0.6cm
{\bf\noindent 1. Introduction}
\vglue 0.2cm

A Rydberg wave packet is formed when a short 
laser pulse excites
a coherent superposition of energy states
\cite{ps,az}.
The initial motion of these localized wave packets
is periodic with a periodicity $T_{\rm cl}$ 
equal to the classical
period of a particle in a keplerian orbit.
However,
after several Kepler cycles, 
the wave packet collapses and a sequence of
fractional and full revivals commences.
The fractional revivals occur prior to the full revival.
They consist of distinct subsidiary waves 
moving with a period that is a fraction of $T_{\rm cl}$
\cite{ps,az,ap}.
These culminate with the formation of a full revival
at time $t_{\rm rev}$
when the wave packet recombines into nearly its original shape.
The fractional revivals have been observed in
time-delayed photoionization
and phase modulation experiments
\cite{expts}.

We begin in the next section by examining
the revival structure and evolution
of hydrogenic Rydberg wave packets
for times much greater than the revival time
\cite{qdsr,sr}.
We show how a quantum-defect theory based on
quantum-mechanical supersymmetry 
\cite{sqdt}
may be used to model
wave packets in alkali-metal atoms,
and we study the effects of quantum defects on the revivals
\cite{qdsr}.
We also describe how this analysis of revivals can
be applied to other quantum systems
\cite{ajp,emp}.
In Sec.\ 3,
we show that the motion of these wave packets
has features characteristic of squeezed states,
and we outline an approach for a squeezed-state description
\cite{rss,esskss}.
Sec.\ 4 examines the revival structure of Rydberg wave
packets moving in the presence of an external electric field
\cite{stark}.
These wave packets are referred to as Stark wave packets.
To lowest order in a perturbative treatment,
Stark wave packets have energies that 
depend on two quantum numbers.
Their evolution is therefore governed by additional
periodicities and revival time scales
\cite{revs2}.
We prove that under certain conditions Stark wave packets can 
exhibit full and fractional revivals.  
We also show that the revivals 
of Stark wave packets have unique features 
arising from the fact that the superposition of 
states can be separated into sums over even and
odd values of the principal quantum number $n$.  
The even and odd states evolve differently,
which results in new types of interference behavior.

\vglue 0.4cm
{\bf\noindent 2. Superrevivals of Rydberg Wave Packets}
\vglue 0.2cm

To date,
experiments that have detected the classical
oscillation of a wave packet in a Coulomb potential
have been performed using purely radial wave packets, 
which consist of a superposition of $n$ states with $l$ fixed,
typically, to a p state.
A wave packet of this type is formed when a single
short laser pulse excites an atom from its ground state
to a superposition of excited states.
Radial wave packets follow the initial classical motion,
but are localized only in the radial coordinate.

The wave function for a hydrogenic radial
wave packet may be expanded in terms of energy eigenstates as
\beq
\Psi ({\vec r},t) = \sum_{n=1}^{\infty} c_n 
\varphi_n ({\vec r}) \exp \left[ -i E_n t \right]
\quad .
\label{wave2}
\eeq
Here,
$E_n = -1/2 n^2$ is the energy in atomic units,
and $\varphi_n ({\vec r})$
represents a radial wave packet,
$\varphi_n ({\vec r}) = \psi_{n,1,0}({\vec r})$,
where $\psi_{nlm}({\vec r})$ is a 
hydrogen eigenstate of 
energy and angular momentum.
The laser is tuned to excite coherently
a superposition of states centered on a central value $\bar n$
of the principal quantum number.
The square of the weighting
coefficients $c_n$ is therefore centered on $\bar n$
with a width characterizing the energy spread.

Expanding the energy in a Taylor series around the centrally
excited value $\bar n$,
we find that the
derivative terms define distinct time scales that depend
on $\bar n$:
\beq
T_{\rm cl} = \fr {2 \pi} {E_{\bar n}^\prime} = 2 \pi {\bar n}^3
\quad ,
\label{tcl}
\eeq
\beq
t_{\rm rev} = \fr {- 2 \pi} {\fr 1 2 E_{\bar n}^{\prime\prime}} 
= \fr {2 {\bar n}} 3 T_{\rm cl}
\quad ,
\label{trev}
\eeq
\beq
t_{\rm sr} = \fr {2 \pi} {\fr 1 6 E_{\bar n}^{\prime\prime\prime}} 
= \fr {3 {\bar n}} 4 t_{\rm rev}
\quad .
\label{tsr}
\eeq
Keeping terms through third order,
and defining the index $k=n-{\bar n}$, 
we may write the wave function as
\beq
\Psi ({\vec r},t) = \sum_{k=-\infty}^{\infty} c_k 
\varphi_k ({\vec r}) \exp \left[ -2 \pi i 
\left( \fr {kt} {T_{\rm cl}} -  \fr {k^2 t} {t_{\rm rev}}
+ \fr {k^3 t} {t_{\rm sr}} \right) \right]
\quad .
\label{psi3rd}
\eeq
The time evolution of the wave packet is controlled by the
interplay between the three time-dependent terms in the phase.
For small times, 
$t \lsim T_{\rm cl}$, 
only the first term in the
phase matters and $\Psi ({\vec r},t)$ is approximately
periodic with period $T_{\rm cl}$.
For times $t \lsim t_{\rm rev}$,
the interference of the first two terms causes the
formation of fractional revivals and full revivals
\cite{ap}.
These occur at times equal to fractions of $t_{\rm rev}$.
The motion of the wave packet at these times is periodic
with periods equal to fractions of the classical
orbital period $T_{\rm cl}$.
The fractional periodicities are evident in plots of
the absolute square of the autocorrelation function
$A(t) = \vev{\Ps(\vec r,0) | \Ps(\vec r,t)}$ as
a function of $t$
\cite{nau,schleich}.

To consider times $t \lsim t_{\rm sr}$,
we must examine the interference between all three
terms in the phase of $\Psi ({\vec r},t)$.
At certain times $t_{\rm frac}$
it is possible to expand
the wave function $\Psi ({\vec r},t)$
of Eq.\ \rf{psi3rd} as a series of subsidiary wave functions.
The idea is to express 
$\Psi ({\vec r},t)$ as a sum of
wave functions $\ps_{\rm cl}$ 
having the same periodicities
and shape as that of
the initial wave function $\Psi ({\vec r},0)$.
We find that at times 
$t_{\rm frac} \approx \fr 1 q t_{\rm sr}$,
where $q$ must be an integer multiple of 3,
the wave packet can be written as a sum of
macroscopically distinct wave packets.
Furthermore,
at these times $t_{\rm frac}$,
we also find that the motion of the wave packet is
periodic with a period 
$T_{\rm frac} \approx \fr 3 q t_{\rm rev}$.
Note that these periodicities are 
on a much greater time scale than
those of the fractional revivals,
and thus a new level of revivals commences
for $t > t_{\rm rev}$.
We also find that at the time
$t_{\rm frac} \approx \fr 1 6 t_{\rm sr}$,
a single wave packet forms that resembles
the initial wave packet more closely than the
full revival does at time $t_{\rm rev}$,
i.e., a superrevival occurs.
In Refs.\ \cite{sr},
we have given theoretical proofs for the
periodicity and occurrence times of the
superrevivals.

One possibility for measuring the
full and fractional superrevivals is
to use a pump-probe method of detection
for radial Rydberg wave packets with  
${\bar n} \approx 45$ -- $50$.
This is experimentally feasible,
provided a delay line of 3 -- 4 nsec
is installed in the apparatus.
This should permit detection of both 
full and fractional superrevivals.

We note that
this analysis of wave-packet revivals
is very general and can be applied to a
variety of quantum systems other than Rydberg atoms
\cite{ajp}.
In fact,
a classification of different revival types is possible
based solely on the form of $E_n$
for quantum systems with discrete energy levels.
For example,
the energy levels of a particle in an infinite square well
depend on $n^2$.
Because of this quadratic dependence,
the time scale $t_{\rm sr}$,
which depends on the third derivative of $E_n$ with
respect to $n$, 
is undefined.
Wave packets in the infinite square well therefore
exhibit perfect full and fractional revivals
and no superrevivals.
A second example, 
in condensed matter physics, 
is of wave packets
formed as superpositions of edge magnetoplasmons
in quantum-Hall devices.
We have shown that these wave packets can exhibit
full and fractional revivals similar to those
of Rydberg wave packets
\cite{emp}.

All of the results described above for hydrogen may be
rederived in the context of
a quantum defect theory based on atomic supersymmetry 
\cite{sqdt}.
The wave functions of this theory form a complete and
orthonormal set with the correct eigenenergies for
an alkali-metal atom.
They are particularly well suited for modeling the
behavior of wave packets in alkali-metal atoms.
The expansion of the energy for
a Rydberg wave packet may be carried out in 
this context with the energies 
$E_{n^\ast} = - 1/{2 n^{\ast 2}}$,
where $n^\ast = n - \de (l)$
and $\de (l)$ is an asymptotic quantum defect
for an alkali-metal atom.
To investigate the effects of a laser detuning,
the Taylor expansion is 
carried out around a noninteger central
value $N^\ast$ that may or may not be on resonance.
For the off-resonance case,
the noninteger part of $N^\ast$ consists of
two parts:  one from the quantum defect and another
from the laser detuning.
In Ref.\ \cite{qdsr},
we have shown
that the effects of the quantum defects
are different from those of the laser detuning.
The time scales governing the evolution of the wave
packet depend on the quantum defects differently
from how they depend on the laser detuning.

\vglue 0.4cm
{\bf\noindent 3. Squeezed-State Description}
\vglue 0.2cm

The initial localization and classical behavior of
radial Rydberg wave packets suggests
that they might be described in terms of 
some kind of coherent state
\cite{cs}.
Indeed,
at an early stage in the development of quantum mechanics,
Schr\"odinger tried unsuccessfully to find nonspreading
wave-packet solutions for a quantum-mechanical
particle in a Coulomb potential
evolving along a classical trajectory
\cite{wavemech}.
Many authors since have discussed this issue,
and it is now known that there are no {\it exact}
coherent states for the Coulomb problem
\cite{brown,mostowski,nieto,mcab,gay,naucs}.

Although the original
Schr\"odinger problem for the Coulomb potential
has no solution,
one can nonetheless obtain minimum-uncertainty
wave packets exhibiting many features of
the corresponding classical motion.
For example,
radial Rydberg wave packets follow the initial classical motion.
However,
they also exhibit distinctive quantum-mechanical features.
In particular,
their uncertainty product in $r$ and $p_r$
oscillates periodically 
as a function of time.
This is a characteristic of a quantum-mechanical squeezed state
\cite{rss}.

To construct a squeezed-state description
appropriate for radial Rydberg wave packets,
we have adopted a procedure used in 
Refs.\ \cite{nieto} 
in the context of the construction
of generalized minimum-uncertainty coherent states.
The idea is to change variables from $r$ and $p_r$
to a new set,
$R$ and $P$,
chosen to have oscillatory dependence on a suitable variable.
The similarities between the ensuing equations and
the usual quantum harmonic oscillator are sufficient
to allow an analytical construction of
our candidate Rydberg wave packets.
Our method generates a three-parameter family
of radial squeezed states
\beq
\psi (r) = 
\fr {(2 \ga_0 )^{2 \al + 3}} {\Ga (2 \al + 3)}
r^{\al} e^{-\ga_0 r} e^{-i \ga_1 r}
\quad .
\label{ras}
\eeq 

To compare our radial squeezed states
with other theory and experiment,
we determine the parameters
$\al$, $\ga_0$, and $\ga_1$ by fixing the form of the
packet at the first pass through the classical apsidal point
by the conditions
$\vev{p_r} = 0$, 
$\vev{r} = r_{\rm out}$,
$\vev{H} = E_{\bar n}$,
where  
$r_{\rm out}$ is 
the outer apsidal point of the orbit
and 
$E_{\bar n} = - 1/2 \bar n^{2}$.
We have shown that these radial squeezed states may
be used as an initial wave function to model the motion of a
wave packet produced by a short laser pulse
\cite{rss}.
The time evolution of the radial squeezed states
exhibits the expected revival structure as well
as the oscillations in the uncertainty product
that are characteristic of a squeezed state.

We have also considered the problem of constructing
squeezed states that are localized in three dimensions
and which follow the classical trajectory of a particle
moving on a keplerian ellipse.
To construct a squeezed state localized 
in the angular coordinates
requires identifying angular operators 
appropriate to a problem with spherical symmetry.
These operators can then be used to obtain a class of squeezed states,
called spherical squeezed states,
which minimize uncertainty products for the angular variables.
Combining the spherical squeezed states 
with radial squeezed states results in a class of
minimum-uncertainty wave packets
that are localized in all three dimensions
and that travel along a keplerian ellipse.
We call these three-dimensional wave packets 
keplerian squeezed states
\cite{esskss}.
The widths of the wave packets oscillate during the motion,
as is characteristic of squeezed states.
The keplerian squeezed states 
maintain their shape for several cycles
before collapsing and undergoing revivals.

Three-dimensional wave packets of this kind 
are of particular interest at present
because experiments using short-pulsed lasers are
attempting to produce Rydberg wave packets that move
along elliptical orbits.
To generate a three-dimensional wave packet 
localized in radial and angular coordinates, 
a superposition of $n$, $l$, and $m$ levels must be created.
This requires the presence of 
additional fields during the excitation process.
One proposal for achieving this
involves using a short electric pulse to
convert an angular state into a localized Rydberg wave
packet moving on a circular orbit
\cite{gns}.
An additional weak electric field could then distort the
orbit into an ellipse of arbitrary eccentricity.
We expect the motion of these wave packets to be 
well described by the keplerian squeezed states.

\vglue 0.4cm
{\bf\noindent 4. Stark Wave Packets}
\vglue 0.2cm

The properties of Rydberg wave packets 
are being investigated
in the presence of electric fields,
which can significantly alter the atomic dynamics.
Here,
we address the question of whether fractional revivals can occur in
these more complicated systems.
Stark wave packets have energies that depend on two quantum numbers
\cite{revs2}.
We prove below that under certain conditions Stark wave
packets can exhibit full and fractional revivals.
Moreover,
we show the existence of new wave-packet behavior that does
not occur for free wave packets
\cite{stark}.

A Stark wave packet is created when
a short laser pulse excites a wave packet
in the presence of a static external electric field.
For a hydrogen atom in a weak electric field, 
the energies in atomic units are 
$E_{n k} = - 1/(2 n^2) + 3nkF/2$,
where $n$ is the principal quantum number,
$k=n_1 - n_2$ with $n_1$ and $n_2$ 
being parabolic quantum numbers,
and $F$ is the magnitude of the electric-field strength.
We use a hydrogenic treatment to describe Stark wave packets.
Wave packets for alkali-metal atoms can be studied using
supersymmetry-based 
quantum defect theory.
The time scales 
describing the evolution of hydrogenic and alkali-metal
wave packets are not equal. 
However,
we expect similar types of revival structure 
for both hydrogen and alkali-metal atoms.

We wish to examine the revival structure of a Stark wave packet 
$\Ps (t)$ formed as a coherent superposition
of states $\ph_{n k}$ with energies $E_{n k}$
depending on two quantum numbers $n$ and $k$.
We write 
$\Psi (t) = \sum_{n, k} c_{n k}
\ph_{n k} \exp \left[ -i E_{n k} t \right]$.
The quantum number $k$ is
even or odd according to whether $n$ is odd or even.
Consider a superposition of Stark states centered around the values
$n = \bar n$ and $k = \bar k = 0$,
and take the quantum number $m$ associated with the third
component of the angular momentum to be zero.
The energy can then be expanded in a Taylor series
around $E_{\bar n \bar k}$.
We introduce the time scales
\beq
T_{\rm cl}^{(n)} 
= \fr {2 \pi} {\left( \fr {\partial E} {\partial n} 
\right)_{\bar n,\bar k}}
= 2 \pi \bar n^3
\quad , \quad\quad\quad
T_{\rm cl}^{(k)} 
= \fr {2 \pi} {2\left( \fr {\partial E} {\partial k} 
\right)_{\bar n,\bar k}}
= \fr {2 \pi} {3 F \bar n}
\quad ,
\label{Tclnk}
\eeq
\beq
t_{\rm rev}^{(n)} = \fr {2 \pi} {\fr 1 2 \left( 
\fr {\partial^2 E} {\partial n^2} 
\right)_{\bar n,\bar k}}
= \fr {4 \pi} 3 \bar n^4
\quad , \quad\quad\quad
t_{\rm rev}^{(nk)} = \fr {2 \pi} {2\left( 
\fr {\partial^2 E} {\partial n \partial k} 
\right)_{\bar n,\bar k}}
= \fr {2 \pi} {3 F}
\quad .
\label{trevnk}
\eeq
No revival time $t_{\rm rev}^{(k)}$ is associated
with the quantum number $k$ since
${\partial^2 E}/{\partial k^2} = 0$.
Substituting these definitions into $\Psi(t)$
and keeping terms to second order yields 
\beq
\Psi (t) = \sum_{n, k} c_{n k}
\ph_{n k} \exp \left[ -2 \pi i
\left(  \fr {(n - \bar n) t} {T_{\rm cl}^{(n)}}
+ \fr {k t} {2 T_{\rm cl}^{(k)}}
+ \fr {(n - \bar n)^2 t} {t_{\rm rev}^{(n)}}
+ \fr {(n - \bar n) k t} {2 t_{\rm rev}^{(nk)}}
\right)
\right]
\quad .
\label{psiexpans}
\eeq

For small times $t$,
the first two terms of the time-dependent phase in 
Eq.\ \rf{psiexpans} dominate.
They represent beating between the two classical periods
$T_{\rm cl}^{(n)}$ and $T_{\rm cl}^{(k)}$.
We say $T_{\rm cl}^{(n)}$ and $T_{\rm cl}^{(k)}$
are commensurate if
$T_{\rm cl}^{(n)} = \fr a b T_{\rm cl}^{(k)}$,
where $a$ and $b$ are relatively prime integers.
If this relation holds,
then the time evolution of $\Ps (t)$ on short time scales
exhibits a period
$T_{\rm cl} = b T_{\rm cl}^{(n)} = a T_{\rm cl}^{(k)}$.

For greater times,
the revival time scales
$t_{\rm rev}^{(n)}$ and
$t_{\rm rev}^{(nk)}$ 
become relevant and modulate the initial behavior,
causing the wave packet to spread and collapse.
We find that the wave packet can undergo full revivals
provided the revival times 
$t_{\rm rev}^{(n)}$ and $t_{\rm rev}^{(nk)}$
are commensurate and obey
$t_{\rm rev}^{(n)} = (r/s) t_{\rm rev}^{(nk)}$,
where $r$ and $s$ are relatively prime integers.
The commensurability of the time scales depends on both
$\bar n$ and $F$.
Restricting the electric-field strength $F$ to 
less than the classical field-ionization threshold 
$F_c = 1/{16 \bar n^4}$
places limits on the ratios $a/b$ and $r/s$.
We find
$a/b < 3/{16}$
and
$r/s < 1/8$.
By tuning $F$ to specific values,
different commensurabilities and types of revival 
structure can be selected.

For fractional revivals to form in Stark wave packets,
the wave function $\Psi(t)$ in Eq.\ \rf{psiexpans}
must be expressible as a sum of distinct subsidiary wave functions.
However,
this can only occur 
at times $t=t_{\rm frac}$
that are simultaneously irreducible rational fractions
of the two revival time scales.
We define the times
$t_{\rm frac} = ({p_1}/{q_1}) t_{\rm rev}^{(n)}
= ({p_{12}}/{q_{12}}) t_{\rm rev}^{(nk)}$.
Here,
the pairs of integers $(p_1,q_1)$
and $(p_{12},q_{12})$ are relatively prime.
To prove that subsidiary wave packets form at the times
$t_{\rm frac}$,
we first rewrite $\Ps(t)$ in Eq.\ \rf{psiexpans}
by shifting $(n - \bar n) \rightarrow n$ and
separating the series into odd and even sums over $n$.
We then let
$k \rightarrow 2k$ in the sum over odd $n$, 
and $k \rightarrow 2k+1$ in the sum over even $n$.
This gives
$\Ps(t) = \Ps_{\rm odd}(t) + \Ps_{\rm even}(t)$,
where $\Ps_{\rm odd}(t)$ and $\Ps_{\rm even}(t)$
are the superpositions of odd-n and even-n states,
respectively.

We define the doubly periodic wave functions
$\ps^{\rm (odd)}_{\rm cl}$ and $\ps^{\rm (even)}_{\rm cl}$ as
\beq
\ps^{\rm (odd/even)}_{\rm cl}(t_1,t_2) = 
\sum_{n \, {\rm odd/even}} \sum_{k} 
c_{nk} \ph_{nk} \,
\exp \left( -2 \pi i
\left(  \fr {n t_1} {T_{\rm cl}^{(n)}}
+ \fr {k t_2} {T_{\rm cl}^{(k)}}
\right)\right)
\quad .
\label{psicloddeven}
\eeq
Next, 
consider the periodicity in $n$ and $k$ for the
higher-order terms in the time-dependent phases of 
$\Ps_{\rm odd}(t)$ and $\Ps_{\rm even}(t)$ at 
$t = t_{\rm frac}$.
These terms are 
\beq
\th^{\rm (odd)}_{nk} = \fr {p_1} {q_1} n^2
- \fr r s \fr {p_1} {q_1} n k 
\quad ,
\label{thetaodd}
\eeq
\beq
\th^{\rm (even)}_{nk} = \fr {p_1} {q_1} n^2
- \fr r s \fr {p_1} {q_1} n k 
- \fr r s \fr {p_1} {q_1} \fr 1 2 n 
\quad .
\label{thetaeven}
\eeq
Here,
$n$ is odd in Eq.\ \rf{thetaodd} 
and even in Eq.\ \rf{thetaeven}.
We seek the minimum periods $l_1$, $l_2$, $l^\prime_1$, 
and $l^\prime_2$ such that
$\th^{\rm (odd)}_{n + l_1,k} = \th^{\rm (odd)}_{nk}$,
$\th^{\rm (odd)}_{n,k + l_2} = \th^{\rm (odd)}_{nk}$,
$\th^{\rm (even)}_{n + l^\prime_1,k} = \th^{\rm (even)}_{nk}$,
and
$\th^{\rm (even)}_{n,k + l^\prime_2} = \th^{\rm (even)}_{nk}$.
These relations yield four conditions 
for the periods $l_1$, $l_2$, $l^\prime_1$, and $l^\prime_2$
in terms of $n$, $k$, and $t_{\rm frac}$.

Since the functions 
$\ps^{\rm (odd)}_{\rm cl}$ and $\ps^{\rm (even)}_{\rm cl}$
with $t$ shifted by appropriate fractions of $T_{\rm cl}^{(n)}$
and $T_{\rm cl}^{(k)}$
have the same periodicities in $n$ and $k$  as
$\th^{\rm (odd)}_{nk}$ and $\th^{\rm (even)}_{nk}$,
respectively,
we may use these functions as a basis for an expansion
of the wave functions $\Ps_{\rm odd}(t)$ and $\Ps_{\rm even}(t)$
at the times $t_{\rm frac}$.
The result is
\bea
\Psi(t) = \sum_{s_1 = 0}^{l_1-1} \sum_{s_2 = 0}^{l_2-1}
a^{\rm (odd)}_{s_1 s_2} 
\ps^{\rm (odd)}_{\rm cl} (t + \fr {s_1} {l_1} T_{\rm cl}^{(n)},
t + \fr {s_2} {l_2} T_{\rm cl}^{(k)})
\qquad\qquad\qquad\qquad\cr
+ \, e^{-i \pi \left( \fr {p_{12} t_{\rm rev}^{(nk)}} 
{q_{12} T_{\rm cl}^{(k)}} \right)} \,
\sum_{s_1 = 0}^{l^\prime_1-1} \sum_{s_2 = 0}^{l^\prime_2-1}
a^{\rm (even)}_{s_1 s_2} 
\ps^{\rm (even)}_{\rm cl} 
(t + \fr {s_1} {l^\prime_1} T_{\rm cl}^{(n)},
t + \fr {s_2} {l^\prime_2} T_{\rm cl}^{(k)})
\quad .
\label{starkfracrev}
\eea

The expansion coefficients $a^{\rm (odd)}_{s_1 s_2}$
and $a^{\rm (even)}_{s_1 s_2}$ are 
\beq
a^{\rm (odd)}_{s_1 s_2} = \fr 1 {l_1 l_2}
\sum_{\ka_1 = 0}^{l_1-1} \sum_{\ka_2 = 0}^{l_2-1}
\exp \left( 2 \pi i \th^{\rm (odd)}_{\ka_1 \ka_2} \right)
\exp \left( 2 \pi i \fr {s_1} {l_1} \ka_1 \right)
\exp \left( 2 \pi i \fr {s_2} {l_2} \ka_2 \right)
\quad ,
\label{assodd}
\eeq
\beq
a^{\rm (even)}_{s_1 s_2} = \fr 1 {l^\prime_1 l^\prime_2}
\sum_{\ka_1 = 0}^{l^\prime_1-1} \sum_{\ka_2 = 0}^{l^\prime_2-1}
\exp \left( 2 \pi i \th^{\rm (even)}_{\ka_1 \ka_2} \right)
\exp \left( 2 \pi i \fr {s_1} {l^\prime_1} \ka_1 \right)
\exp \left( 2 \pi i \fr {s_2} {l^\prime_2} \ka_2 \right)
\quad ,
\label{asseven}
\eeq
When these expressions are substituted into
Eq.\ \rf{starkfracrev} and the definitions 
\rf{psicloddeven} are used,
Eq.\ \rf{starkfracrev} reduces to the form given in
Eq.\ \rf{psiexpans}.
This completes the proof of the formation of
fractional revivals for Stark wave packets.

As an illustrative example,
consider the case $\bar n = 24$,
and set $t_{\rm rev}^{(n)}/t_{\rm rev}^{(nk)} = r/s = 1/12$
by tuning the electric-field strength to $F \simeq 645.8$
volts/cm.
For this example,
$t_{\rm rev} = t_{\rm rev}^{(nk)} = 12 t_{\rm rev}^{(n)}$.
We find at $t \approx t_{\rm rev}$,
\beq
\Ps(t) \approx \ps^{\rm (odd)}_{\rm cl}(t,t) +
\ps^{\rm (even)}_{\rm cl}(t,t)
\quad .
\label{dieci}
\eeq
These sums are in phase and combine as a single total
wave packet,
producing the full revival at $t_{\rm rev}$.

At $t = t_{\rm rev}/2$,
however,
we find a time phase between $\Ps_{\rm odd}(t)$ and
$\Ps_{\rm even}(t)$,
with the full wave function reducing to
\beq
\Ps(t) \approx \ps^{\rm (odd)}_{\rm cl}
(t,t + \frac 1 2 T_{\rm cl}^{(k)})
 + \ps^{\rm (even)}_{\rm cl}(t + \frac 1 4 T_{\rm cl}^{(n)},t)
\quad .
\label{undieci}
\eeq
We see that this fractional revival
consists of two subsidiary wave functions out of phase
with each other.

Due to the unique behavior of the quantum number $k$,
the functions $\ps^{\rm odd}_{\rm cl}$ and
$\ps^{\rm even}_{\rm cl}$ exhibit additional
time behavior depending on $\half T_{\rm cl}^{\rm (n)}$.
The functions $\ps^{\rm odd}_{\rm cl}$ are antiperiodic
with period $\half T_{\rm cl}^{\rm (n)}$,
while $\ps^{\rm even}_{\rm cl}$ is periodic in
$\half T_{\rm cl}^{\rm (n)}$.
They obey the relations,
\beq
\ps^{\rm (odd)}_{\rm cl}(t + \half T_{\rm cl}^{(n)},t)
= - \ps^{\rm (odd)}_{\rm cl}(t,t)
\quad ,
\label{per1}
\eeq
\beq
\ps^{\rm (even)}_{\rm cl}(t + \half T_{\rm cl}^{(n)},t)
= \ps^{\rm (even)}_{\rm cl}(t,t)
\quad .
\label{per2}
\eeq
It is this additional behavior with period 
$\half T_{\rm cl}^{\rm (n)}$ that causes Stark wave packets to
have unconventional revival structure.
In Ref.\ \cite{stark},
we proved that at the $t = t_{\rm rev}/2$
fractional revival,
the antiperiodic behavior of 
$\ps^{\rm (odd)}_{\rm cl}$ causes nodes to
appear in the autocorrelation function
with a periodicity equal to $\half T_{\rm cl}^{(n)}$.
However, 
no additional nodes or periodicities appear for the even states.
This additional interference behavior has no analogue for the
case of free Rydberg wave packets.

\vglue 0.4cm

This work is supported in part by the National
Science Foundation under grant number PHY-9503756.

\vglue 0.4cm
{\bf\noindent 5. References}
\vglue 0.2cm

\end{document}